\documentclass[12pt]{article}
\usepackage{psfig}
\usepackage{epsf}

\begin{document}

\title{ {\bf Molecular dynamics study of melting of a bcc metal-vanadium II :
thermodynamic melting}  }% Force line breaks with \\

\author{ V.Sorkin, E. Polturak and Joan Adler }
\date{ Dept. of Physics,~Technion Institute of Technology,~32000 Haifa,~Israel}
\date{5 May 2003 }
\maketitle

\begin{abstract}
We present molecular dynamics simulations of the thermodynamic
melting transition of a bcc metal, vanadium using the
Finnis-Sinclair potential. We studied the structural, transport
and energetic properties of slabs made of 27 atomic layers with a
free surface. We investigated premelting phenomena at the
low-index surfaces of vanadium; V(111), V(001), and V(011),
finding that as the temperature increases, the V(111)
surface disorders first, then the V(100) surface, while the V(110)
surface remains stable up to the melting temperature. Also, as the
temperature increases, the disorder spreads from the surface layer
into the bulk, establishing a thin quasiliquid film in the surface
region. We conclude that the hierarchy of premelting phenomena is
inversely proportional to the surface atomic density, being most
pronounced for the V(111) surface which has the lowest surface
density.
\end{abstract}

\section{\label{sec:level1}  Introduction}

Theories of melting \cite{Ubb,Lind,Dash,Ida} can be separated into
two classes. The first one describes the {\it mechanical melting}
of a homogeneous solid resulting from lattice instability~\cite{
Born, Tallon, Wolf} and/or the spontaneous generation of thermal
defects \cite{ Granato,Weber,Cahn2, Kanigel}. The second class
treats the {\it thermodynamic melting} of heterogeneous solids,
which begins at extrinsic defects such as a free surface or an
internal interface (grain boundaries, voids, etc) \cite{Dash, Dash1, Cahn1,Frenken,
Phill, Veen, Traynov, Barnett}. Both types of
melting have been studied extensively for the case of fcc metals.
In order to determine whether the mechanism of homogeneous melting
differs if the lattice symmetry is changed, in our first
paper \cite{Sorkin}, we extended these studies to a bcc metal,
vanadium. We used molecular dynamics simulations to study the
mechanical melting transition in a bulk system, and found that the
melting transition occurs uniformly throughout the solid at the
melting point. The shear elastic instability leading to this
melting of vanadium occurred once the molar volume reached a
critical value, whether reached by heating the solid or
by adding defects at a constant temperature \cite{Sorkin}. The
temperature at which the mechanical melting transition takes place
in our model is $T_s=2500\pm 10$ K, which is above the melting
temperature measured experimentally, $T_m=2183 $ K. Thus, the
mechanical melting transition can be observed only if melting at
surfaces is artificially suppressed, since in the laboratory it is
preempted by the thermodynamic melting transition which begins at
a free surface.

Theoretical aspects of melting beginning at the surface have been
studied by phenomenological
theories \cite{Veen,Lipowsky1,Lipowsky2,Lipowsky3,Tomagnini},
lattice-gas models \cite{Traynov}, and density functional theory
\cite{Lowen}. Microscopic descriptions of static and dynamic
properties of  melting phenomena beginning at a surface emerged
from computer simulations which employ many-body interaction
potentials derived from the
effective-medium~\cite{Hakkinen,Hakkinen1,Beaudet} and
embedded-atom theories \cite{Chen,Chen1,Barnett,Stoltze,Cox}, as
well as pairwise interatomic interactions in the form of
Lennard-Jones potentials \cite{Broughton,Broughton1}. Melting at
the surface is anisotropic, meaning that certain surfaces of fcc
metals (Pb(110), Al(110)) exhibit premelting,\cite{Veen} while
this phenomenon is not observed for the close-packed surfaces
(Pb(111), Al(111)).  Theoretical aspects of this anisotropy were
discussed by Trayanov and Tosatti \cite{Traynov}.

This paper is organized as follows: details of the molecular-dynamics simulations are presented in
Sec. II. Sec. III contains most of the results of this study,
including the thermodynamic melting temperature and several
structural, energetic, and transport properties of the surface.
Finally, in Sec. IV, we summarize our results.

\section{\label{sec:level1}  Simulation details  }

We model the melting of vanadium with a free surface using
molecular dynamics (MD) simulations~\cite{Rapaport, Allen} in a
canonical ensemble.  The many-body interaction potential used in
this study was developed by Finnis and Sinclair \cite{Finnis} (FS).
The computational aspects of the algorithm are described in detail
in our previous article \cite{Sorkin}.

\begin{figure}[t]
\centerline{\epsfxsize=7.0cm \epsfbox{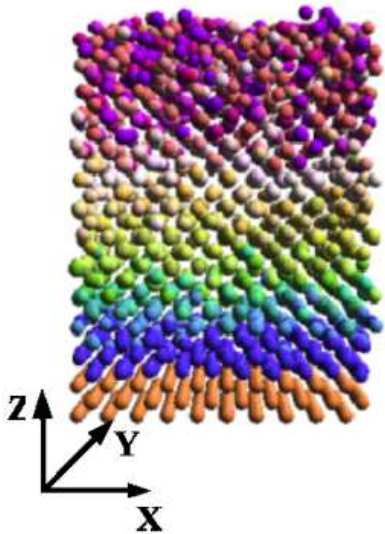}}
\caption{\label{pbc}  Geometry of the sample (V(001) surface).}
\end{figure}

Ideally, a crystal with a surface is a semi-infinite system. We
modeled this system as a thick slab. This slab is shown in Fig.
\ref{pbc}. The bottom 3 atomic layers have the positions of the
atoms fixed in order to mimic the effect of the presence of the
infinite bulk of the crystal. Due to the relatively short range of
the repulsive part of the modified FS potential, 3 fixed layers
are sufficient to represent the static substrate. On top of those
3 layers there are 24 layers in which the atoms are free to move.
Periodic boundary conditions are used along the in-plane (x and y)
directions, while the top boundary of the slab, along the z
direction, is free. In Figure \ref{pbc} we show a sample
where the temperature is high enough so that the layers near the free
surface are already disordered.

\begin{figure}[t]
\centerline{\epsfxsize=15.0cm \epsfbox{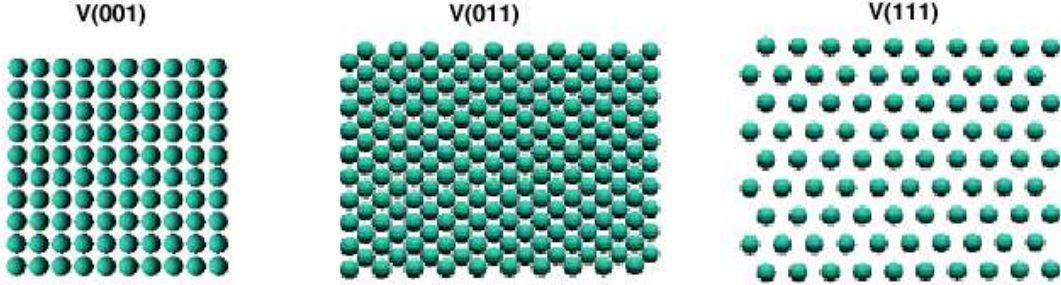}}
\caption{\label{Va111} Geometry of the low-index faces of vanadium
(top view). The drawings are to scale.}
\end{figure}

Three different samples with various low-index surfaces were
constructed: V(001), V(011) and V(111), whose surface layers are
shown in Fig. \ref{Va111}. The V(001) and V(111) samples contain
2700 atoms, initially arranged as a perfect bcc crystal of
10x10x27 unit cells (100 atoms in a layer). The V(011) sample
contains 2646 atoms arranged as  7x7x27 unit cells (98 atoms in a
layer).

Each simulation starts from a low-temperature solid. The lattice
constant of the frozen layers is assigned a value appropriate to
the bulk\cite{Sorkin}. As the temperature is raised, this value is
adjusted to reflect thermal expansion as obtained from bulk
simulations\cite{Sorkin}. We found that $\sim 10^{5}$ integration
time steps were sufficient in order to reach thermal equilibrium
after a temperature change (a time step $dt~=~1.05\times10^{-15}$
sec). An equilibrium state is considered to be achieved when there
is no significant temporal variation (beyond the statistical
fluctuations) of the total energy, layer occupation number,
structure order parameters, self diffusion coefficients, and a
uniform profile of kinetic temperature across the sample is
observed. Thereafter the various structural and transport
properties of the system are calculated, accumulated and averaged
over a long period of $10^{7}$ MD steps. Throughout this study,
interactive visualization (the AViz program~\cite{Adler}) was
implemented, to observe sample disorder and layer mixing (see
Fig.\ref{pbc}). A discussion of this visualization can be found
in Ref.~\cite{Adler2}.

\section{\label{sec:level1} Premelting effects on surfaces}
\subsection{\label{sec:level1}  Thermodynamic melting temperature }

In order to investigate the phenomenon of surface disorder and
premelting, we first determined the thermodynamic melting point
$T_m$ of vanadium described by the modified FS potential. $T_m$
was obtained by means of the method proposed by  Lutsko et
al.~\cite{Lutsko}. In this way, we found $T_m=2220 \pm 10 K$ for
the V(111) sample, which has the lowest surface packing density. This
value is close to the experimental value ~\cite{Handbook} of
$T_m=2183 K$. For the other surfaces, the value of $T_m$ comes out
slightly higher, but within the uncertainty limits of the
simulation. The difference could be due to superheating effects on
the solid-liquid interface just above the thermodynamical melting
point $T_m$. The largest difference is found for the close packed
V(011) surface, with $T_m \approx 2240 \pm 10 ~K$. This dependence
of the thermodynamic melting point on the crystalline plane
forming the surface was also observed in MD simulations of fcc
metals ~\cite{Hakkinen, Beaudet,Chen,Chen1,Kanigel}.

\subsection{\label{sec:level1} Surface thermal expansion and amplitude of atomic vibration}

The interlayer relaxation and surface thermal expansion was calculated
using the difference between the average $z$ coordinate of the $i$th
and $i+1$th layers:
\begin{equation}
d_{i,i+1}=\left<  \frac{1}{n_{i+1}} \sum_{j \in i+1} z_j-
\frac{1}{n_{i}} \sum_{j \in i} z_j  \right >
\end{equation}
where $z_i$ is the $z$-coordinate of the atom $i$, $n_i$ is the
instantaneous number of atoms in the layer $i$, and the angular
brackets denote a time average. The distances between the
neighboring layers could be determined up to the temperatures
where surface premelting effects blur the distinction between
individual layers, although some structure remains visible in the
local density profiles (see below).

\begin{figure}[p]
\centerline{\epsfxsize=9.0cm \epsfbox{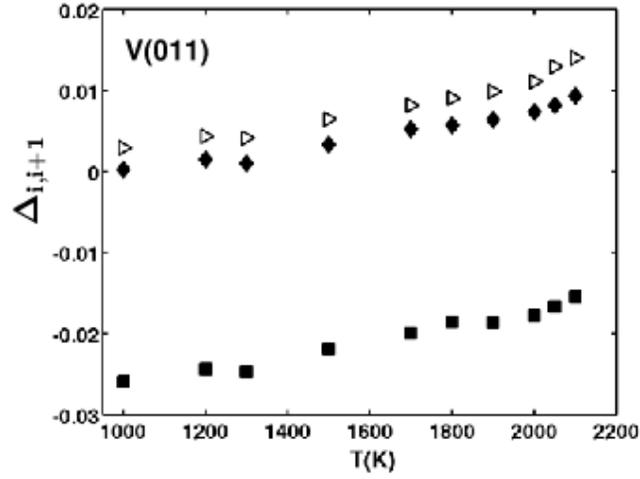}}
\caption{\label{nnDist}  Plot of $\Delta_{i,i+1}$ for the first (squares), second (triangles)
 and third (diamonds) surface layers as a function of temperature for the V(011) surface.}
\end{figure}
\begin{figure}[p]
\centerline{\epsfxsize=9.0cm \epsfbox{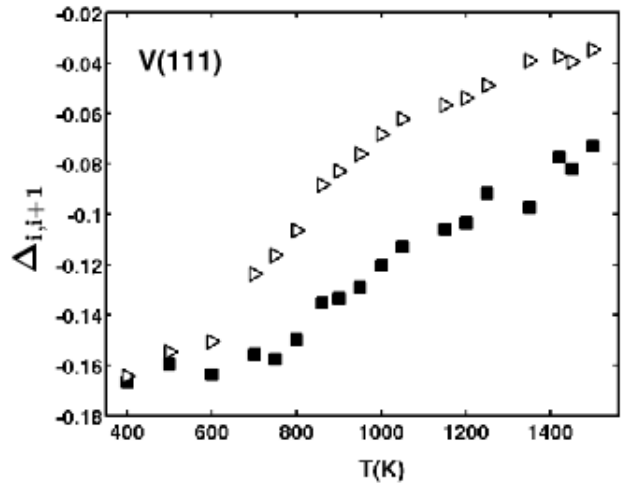}}
\caption{\label{nnnDist}  Plot of $\Delta_{i,i+1}$ for the first (squares) and second (triangles)
 surface layers as a function of temperature for the V(111) surface.}
\end{figure}

At low temperatures, the first layer exhibits an inward
relaxation, i.e. $\Delta_{1,2} < 0 $, where we define
\begin{equation}
\Delta_{i,i+1}=\frac{d_{i,i+1}-d_{bulk}}{d_{bulk}}
\end{equation}
where $d_{i,i+1}$ is the distance in the $z$-direction between the
$i$ and $i+1$ surface layers, and  $d_{bulk}$ is the distance between
two neighboring layers in the bulk. In Figures~\ref{nnDist} and
\ref{nnnDist} $\Delta_{i,i+1}$ is plotted versus temperature
for the two of the low-index faces.

\begin{figure}[h]
\centerline{\epsfxsize=10.0cm \epsfbox{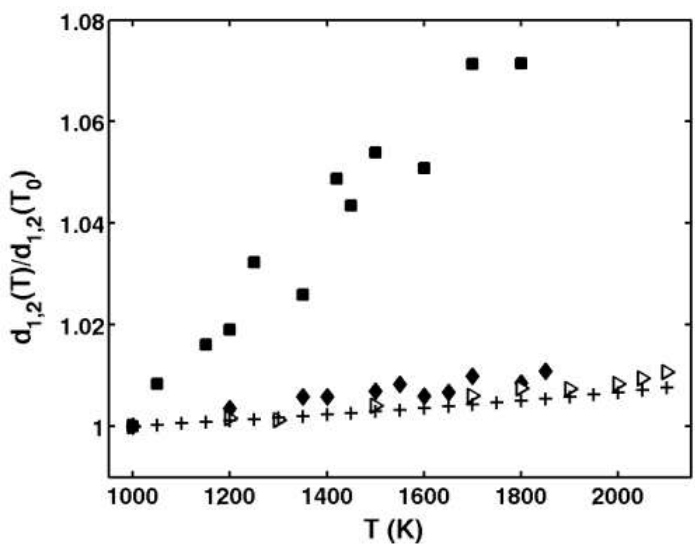}}
\caption{\label{nDist1} Normalized interlayer distances between
the first and second layers, $d_{1,2}(T)/d_{1,2}(T_0)$ at $T_0=1000K$
for the $V(111)$ (squares), $V(001)$ (diamonds),  $V(011)$ (triangles) surfaces
and the bulk (crosses). Note the anomalous thermal
expansion of the $V(111)$ surface.}
\end{figure}

\begin{figure}[h]
\centerline{\epsfxsize=10.0cm \epsfbox{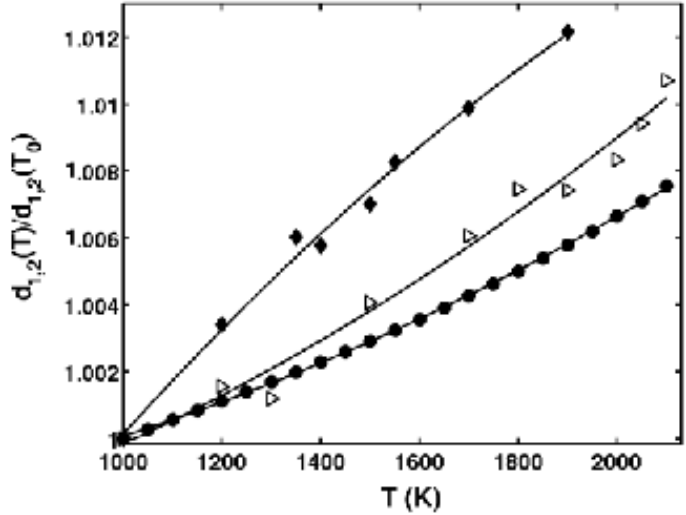}}
\caption{\label{nDist2} Normalized interlayer distances between
the first and second layers, $d_{1,2}(T)/d_{1,2}(T_0)$ at
$T_0=1000K$ for the $V(001)$ (diamonds) and $V(011)$ (triangles) and
the bulk (circles). The solid lines are drawn to guide the eye.}
\end{figure}

The effect of the different geometry of the various surfaces is
reflected in the thermal expansion, i.e. a surface with a lower
atomic density expands more than a surface which is close packed.
As shown in Fig.~\ref{nDist1}, the thermal expansion of the V(111) surface
layers is significantly larger than the thermal expansion of other surfaces
and fixed layers (bulk). In Fig.~\ref{nDist2} we plot thermal expansion of
surface layers (for the V(001) and V(011) slabs) and fixed layers on an enlarged scale.  
The observed ``anomalous'' thermal expansion of
these surface layers is a direct consequence of enhanced vibration
at the surface, where atoms probe the more anharmonic region of
the interatomic potential. 
 Anomalous thermal expansion was first
observed in experiments on Pb(110) by Frenken~\cite{Frenken}.

Atomic vibration properties at the different faces of vanadium are
important for illustrating the onset of anharmonicity at high
temperatures. We calculate the mean square vibration amplitude
(MSVA), $<u^2_{\alpha}> $, around the equilibrium position in the
first surface layer as:
\begin{equation}
<u^2_{\alpha} >=\frac{1}{n_1}\sum_{i=1}^{n_1} \left < \left[ \vec r_{i,\alpha(t)}- <\vec r_{i,\alpha(t)}>  \right]^2 \right>
\end{equation}
where $\alpha=x,y,z$ denotes the spatial direction of motion,
$n_1$ is the instantaneous number of atoms in the first surface
layer, and the angular brackets represent an average over time.

\begin{figure}[p]
\centerline{\epsfxsize=9.0cm \epsfbox{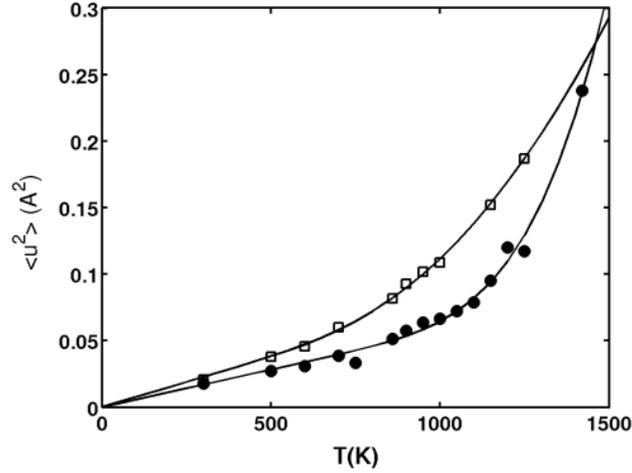}}
\caption{\label{uxz_111} Mean-square amplitudes of vibration  $<u_z^2>$
(filled circles) and $<u_x^2>$ (squares) for the first layer of the V(111) surface
as a function of temperature. The solid lines are a fit to the data. }
\end{figure}

In a harmonic system, $<u^2_{\alpha}
> $ increases linearly with temperature, and therefore deviations from linearity
are a measure of anharmonicity. The mean square amplitudes of
vibration shown in Fig. \ref{uxz_111} for the V(111) surface
demonstrate substantial anharmonicity at elevated temperatures.
 Figure \ref{uxz_111} also shows that there is a noticeable
difference between the in-plane MSVA, $<u^2_{x} > $, and
out-of-plane one, $<u^2_{z}
> $ (we found that $<u^2_{x} > \simeq <u^2_{y}
>$).

\begin{figure}[p]
\centerline{\epsfxsize=9.0cm \epsfbox{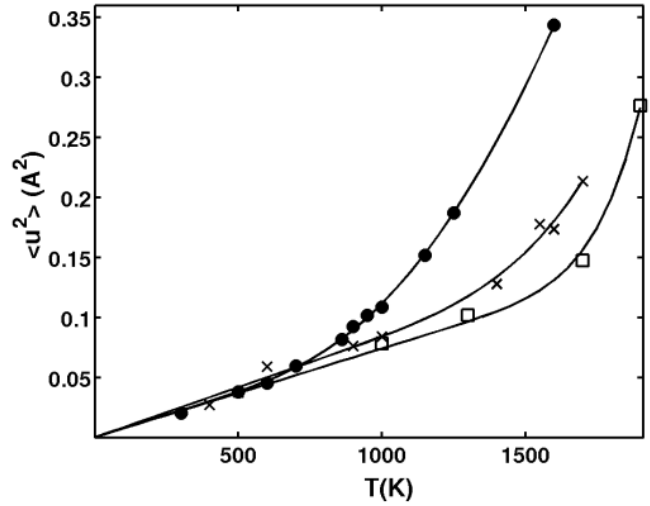}}
\caption{\label{u2} The calculated in-plane compoments of mean square
amplitudes of vibration  $<u_y^2>$ for the first surface layer of V(111) (filled circles),
V(001) (croses) and V(011) (squares) as a function of temperature.
The solid lines are a fit to the data.  }
\end{figure}

 In-plane MSVA's are larger than
out-of-plane amplitudes also for V(001) and V(011) surfaces. One
reason for this anisotropy could be the different nearest
neighbour distance along the $x$ and $z$ directions caused by the
inward relaxation of the surface layer, leading to a net restoring
force that is higher in the direction perpendicular to the surface.
 The in-plane components of the mean square amplitudes of
vibration for the various faces of vanadium are shown in Fig.
\ref{u2}. As may be seen from the figure the in-plane MSVA is
largest for the V(111), and smallest for the V(011)surface. Atoms
which belong to the loose-packed surface V(111) are less tightly
bound than atoms of the close-packed  V(011) face, and therefore
vibrate with larger amplitudes.

\subsection{\label{sec:level1} Layer density profiles}

\begin{figure}[h]
\centerline{\epsfxsize=10.0cm \epsfbox{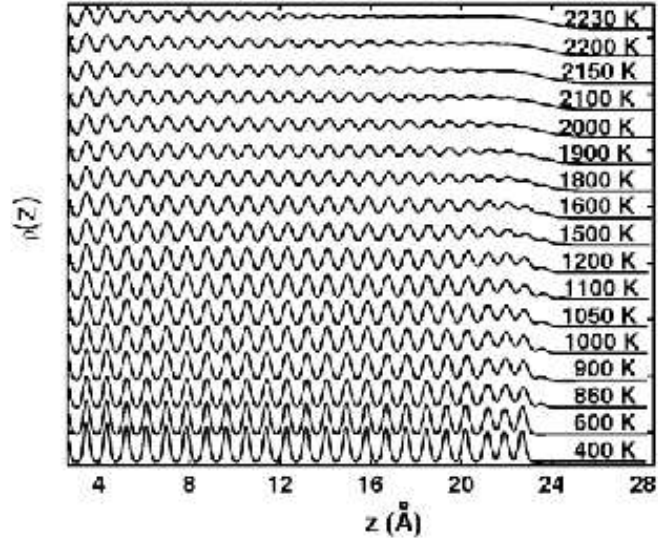}}
\caption{\label{dens111} Density profile across the V(111) slab
along the $z$ direction at various temperatures.}
\end{figure}

\begin{figure}[p]
\centerline{\epsfxsize=10.0cm \epsfbox{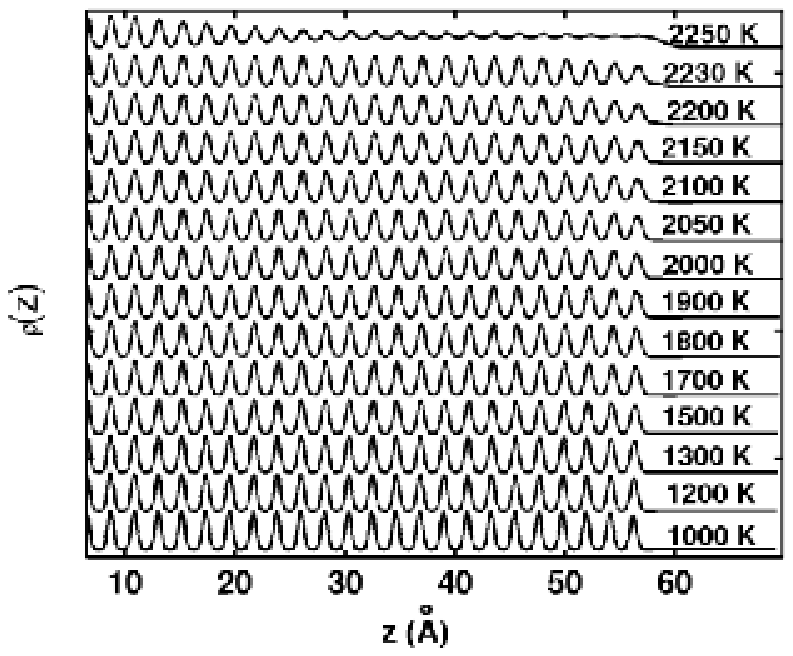}}
\caption{\label{dens001}Density profile across the V(011) slab
along the $z$ direction at various temperatures. }
\end{figure}

\begin{figure}[p]
\centerline{\epsfxsize=10.0cm \epsfbox{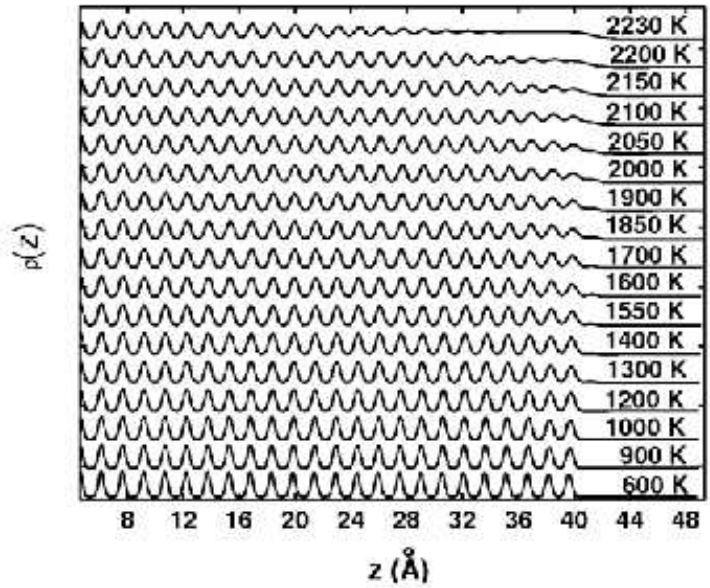}}
\caption{\label{dens011}Density profile across the V(001) slab
along the $z$ direction  at various temperatures.}
\end{figure}

In order to display the structure of the sample along the $z$
direction, perpendicular to the surface, it is convenient to use
the density $\rho(z)$, defined so that $\rho(z)dz$ is the number
of atoms in a slice of thickness $dz$ at $z$. We represent $\rho(z)$ 
by a continuous function defined according to Chen et
al. \cite{Chen,Chen1}
\begin{equation}
\rho(z)=\left<\frac{1}{\sqrt{2\pi \Delta z}} \sum_{i} \exp(-\frac{(z-z_i)^2}{2\Delta z})\right >
\end{equation}
where $z_i$ is the $z$ coordinate of atom $i$, with $z=0$ set at the
bottom of the first layer which is not fixed, and the angular
brackets indicate a time average. We use $\Delta z~
=~0.1a_0/2\sqrt3 $, where $a_0$ is the bulk lattice parameter at a
given temperature. For a system in equilibrium, $\rho(z)$ can be
obtained by an average over many different configurations.
Plots of $\rho(z)$ at different temperatures are given in
Fig.~\ref{dens111}  for the V(111) slab, in Fig.~\ref{dens011} for
the V(011) slab and in Fig.~\ref{dens001} for the V(001) slab.

Premelting effects at the crystal surface can be examined by
monitoring the layer-by-layer modulation of the density profile of
the system, $\rho(z)$, at various temperatures up to the melting
point $T_m$. At low temperatures $\rho(z)$ consists of a series of
well resolved peaks. The atoms are packed in the layers with
constant density in each layer and virtually no atoms in between
these layers. As the temperature increases the effective width of
each layer becomes broader due to the enhanced atomic vibration.
The position of the density peaks moves to larger values of $z$
due to the thermal expansion. At higher temperatures the atomic
vibration becomes so large, especially in the topmost layers, that
the minima of $\rho(z)$ in between two layers rise to non-zero
values. In addition, disorder sets in, with atomic migration
taking place between the layers. Comparing
Figs.~\ref{dens111}-\ref{dens011}, one can see that the V(111)
surface crosses over to a state of premelting first, than V(001),
while on the V(011) surface premelting effects are very small.

Above some temperature ($T_{pm} \simeq~ 1000~K$ for the V(111) surface,
$T_{pm}\simeq~ 1800~K$ for the V(001) surface, $T_{pm}\simeq~ 2200~K$ for
the V(011) surface) the density of the topmost layer becomes slightly
lower than that of the underlying layers. This loss of density is
compensated by the appearance of an adlayer. The adlayer
appears  first on the least packed surface V(111), and then at a higher
temperature on the V(001) surface. The close packed face V(011)
develops an adlayer at temperatures very close to $T_m$. This
hierarchy indicates that the formation energy of structural
defects (vacancy-adatom pairs) is different on these surfaces.
That hierarchy was also observed in the investigation of the surface premelting of
fcc metals (Al\cite{Stoltze,Stoltze1},
Ni~\cite{Chen,Chen1,Cox,Beaudet},
 and Cu~\cite{Chen,Barnett}), where it was found that an adlayer
appears first on the least packed
(011) surface, then at higher temperature on the (001) face, and
finally on the close packed (111) surface at a temperature close
to $T_m$. As the temperature increases towards $T_m$ the
distinction between the surface layers becomes blurred, which is
consistent with the topmost layers converting into a quasiliquid.

\begin{figure}[p]
\centerline{\epsfxsize=8.0cm \epsfbox{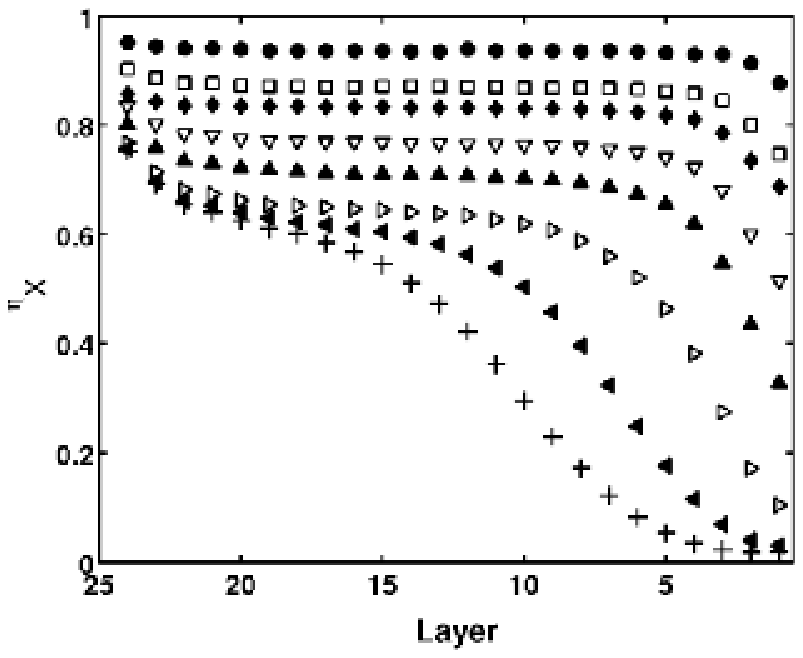}}
\caption{\label{ord111} In plane structure factor, $\eta_x$, of
the V(111) slab along the $x$ direction vs layer number
 at several temperatures: $T=400 K$ (circles), $T=860 K$ (squares), $T=1100 K$ (diamonds),
$T=1500 K$ (triangles down), $T=1800 K$ (triangles up), $T=2100 K$ (triangles left), $T=2200 K$ (triangles right),
and $T=2230 K$ (crosses). }
\end{figure}
\begin{figure}[p]
\centerline{\epsfxsize=8.0cm \epsfbox{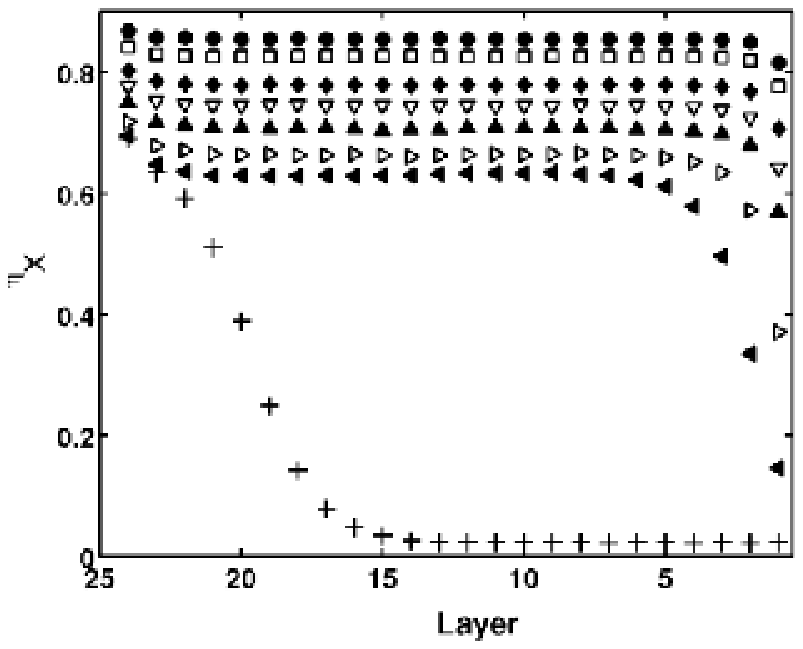}}
\caption{\label{ord011} In-plane structure factor, $\eta_x$, of
the V(011) slab calculated along $x$ direction vs layer number
 at several temperatures: $T=1000 K$ (circles), $T=1200 K$ (squares), $T=1500 K$ (diamonds),
$T=1700 K$ (triangles down), $T=1900 K$ (triangles up), $T=2100 K$ (triangles left), $T=2230 K$ (triangles right),
and $T=2250 K$ (crosses).  }
\end{figure}

\subsection{\label{sec:level1} Layer structure factors}

The preceding section dealt with the onset of disorder along the
$z$ direction, perpendicular to the surface. In order to monitor
the onset of in-plane disorder in each atomic layer,
it is useful to define a structure factor (order
parameter), $\eta_{l,\alpha}$, through a Fourier transform of the atomic density which
is calculated separately for each layer
\begin{equation}
\eta_{l,\alpha}=\left< \frac{1}{n_l^2} \left | \sum_{j \in l } \exp{(i\vec g_{\alpha} \vec r_j)} \right |^2  \right >
\end{equation}
where $\alpha = x,y $ and $ \vec g_\alpha= {2\pi}{/a_\alpha \vec
e_\alpha}$ is a set of reciprocal lattice vectors. $ a_\alpha $ is
the distance between nearest neighbors, and $n_l$ is the
instantaneous number of atoms in the layer $l$. The sum extends
over the particles in the layer $l$, and the angular brackets
denote averaging over time.

For an ordered crystalline surface the structure factor is a unity
at zero temperature. Enhanced vibration and formation of point
defects lead to a decrease of the structure factor with increasing
temperature. This is illustrated in Figs.~\ref{ord111} and \ref{ord011}
 where the structure factor along the $x$
direction is plotted vs. layer number for the V(111) and the
V(011) slab, respectively. It is seen that these effects are most
pronounced in the surface region. Vacancies do not  directly
affect the order parameter, since a normalization procedure is
employed during each measurement by using the instantaneous layer
occupation, $n_l$, of a layer. Nevertheless, vacancies have an
indirect effect on the order parameter by introducing a localized
lattice distortion.

As Figs.~\ref{ord111} and \ref{ord011} show, for the V(111) slab
we found a continuous decrease of the in-plane order parameter. In
contrast, the close packed V(011) sample exhibits a relatively
abrupt decrease of the structure factor within less than $\sim$
20K of the melting temperature.

\subsection{\label{sec:level1} Radial distribution function }
The formation of a quasiliquid film can be analyzed by using a plane radial distribution function defined as
\begin{equation}
p_l(r_{||})=\left< \frac{1}{n_l} \sum_{i,j \in l } \frac{\delta (r_{ij,||}-r_{||})}{2 \pi r_{||}} \right >
\end{equation}
where $r_{ij,||}$ is the component of  the $\vec r_{i}-\vec r_{j} $ parallel to the surface
plane,  $n_l$ is the instantaneous number of atoms in layer $l$, the sum extends over all particles in layer $l$,
and the angular brackets denote averaging over time.

\begin{figure}[h]
\centerline{\epsfxsize=10.0cm \epsfbox{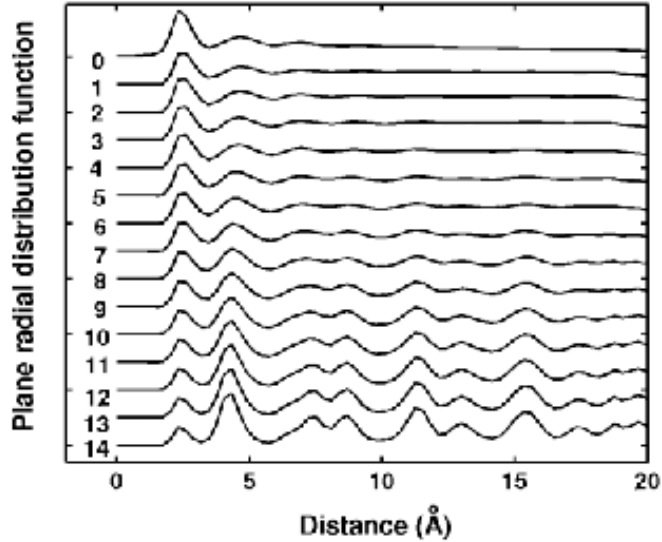}}
\caption{\label{rdf_2200k} Two-dimensional radial distribution
function $p(r_{||})$ of the top surface layers of  V(111) at
temperature T=2230K. The layer n=0 corresponds to the adlayer,
 n=1 to the first layer, n=2  to the second one, etc.}
\end{figure}

\begin{figure}[p]
\centerline{\epsfxsize=8.5cm \epsfbox{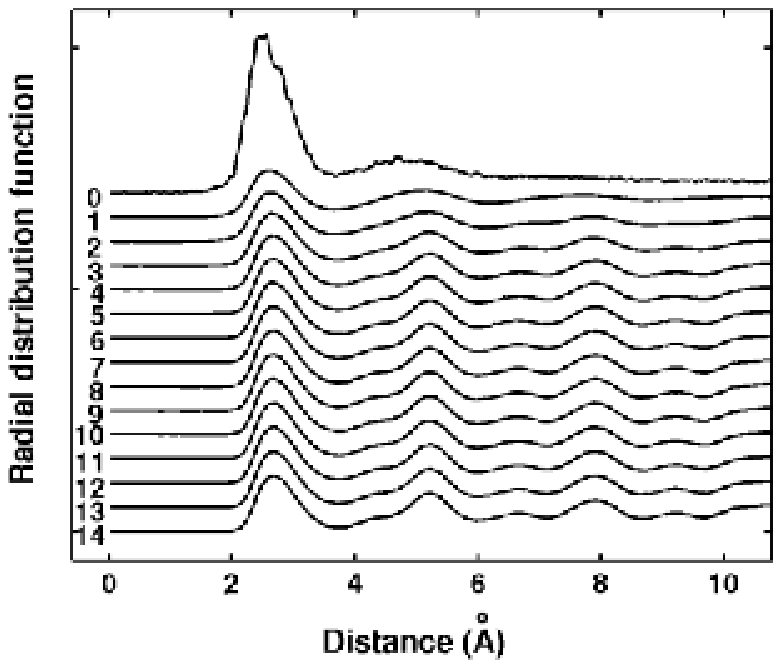}}
\caption{\label{rdf_2200k_1} Two-dimensional radial distribution
function $p(r_{||})$ of the surface layers of V(011) at
temperature T=2230K. The layer n=0 corresponds to the adlayer, n=1
to the first layer, n=2 to the second one, etc.}
\end{figure}

The two-dimensional radial distribution function, $p(r_{||})$, for
several layers of the V(111) and V(011) slabs very close to $T_m$
is shown in Figs.~\ref{rdf_2200k} and~\ref{rdf_2200k_1}. As
seen from the figures the intra-layer structure in these layers
changes gradually from crystalline to liquid-like as one
approaches the surface. Particularly noticeable is the
disappearance of the crystalline features in $p(r_{||})$ that
correspond to the second, third and other nearest neighbors. In
addition to the heights of the peaks, the area under the
$p(r_{||})$ curve changes, which reflects the change in the
density across the solid-liquid interface.
\begin{figure}[p]
\centerline{\epsfxsize=8.5cm \epsfbox{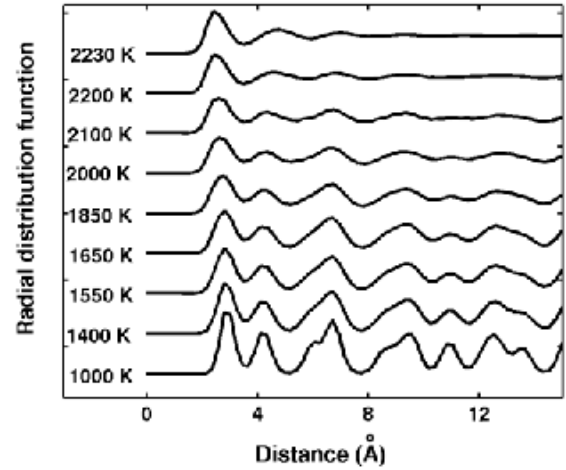}}
\caption{\label{rdf_1_001}Two-dimensional radial distribution function $p(r_{||})$
of the first layer of V(001) at various temperatures.}
\end{figure}

We note that within the adlayer, the probability of finding
particles with separation beyond the first-neighbor shell is
relatively small, indicating a tendency for clustering  which
persists, though to a smaller extent, even to $T \simeq T_m$.
The plane pair-correlation functions for the first layer of the
V(001)  at elevated temperatures are shown in
Fig.~\ref{rdf_1_001}. It is clear that the crystalline structure
vanishes gradually and the layer becomes quasiliquid  as the
melting point $T_m$ is approached.

We conclude from the analysis of the plane radial distribution
functions  and density profiles that surface premelting begins
first on the least packed surface V(111), while changes only
appear on the closed packed V(011) surface at temperatures which
are very close to the melting point.

\subsection{\label{sec:level1} Layer occupation and point defects}

As seen in the density profiles (Figs.~\ref{dens111}-ref{dens001}), the formation of an adlayer on the least packed
surface V(111) begins at around $T~ \sim~ 800$ K. At these
temperatures, the appearance of an adlayer involves generation of
vacancies only in the first surface layer, while at higher
temperatures ($T \ge 1600 K $) vacancies in the underlying layers
(the second and third layers) begin to appear via promotion of
atoms to vacant sites in the first surface layer. Atom migration
from the deeper layers into the surface layers increases
significantly as the temperature approaches $T_m$. In contrast to
the V(111) sample, adlayer formation and generation of
adatom-vacancy pairs at the the V(001) surface becomes observable
only above $ T~ \sim ~1400K$ (in practice, all adatoms come from
the first layer).

The same effect, namely the appearance of an adlayer at different
surfaces at successively higher temperatures was observed in
computer simulations of surface premelting of fcc metals
\cite{Hakkinen,Chen,Chen1,Barnett,Stoltze}, where the lowest
density (110) surface of fcc metals begins to disorder first,
while the close packed (111) surface preserves its ordered
crystalline structure almost up to $T_m$.

\begin{figure}[h]
\centerline{\epsfxsize=10.0cm \epsfbox{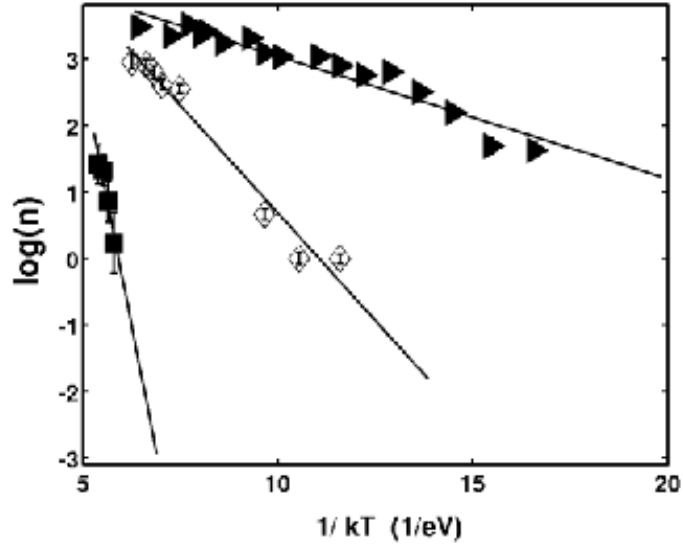}}
\caption{\label{adlocc} The natural logarithm of the equilibrium
adatom concentration, $n$, as a function of $1/k_bT$  on the
$V(001)$, $V(011)$ and $V(111)$ faces (denoted by diamonds,
triangles and squares, respectively.) The dashed lines are linear
fits to the data.}
\end{figure}

Knowledge of the equilibrium averaged number of atoms in the
adlayer allows us to estimate $E_s $, the formation energy of a
surface defect (adatom-vacancy pair) according to the relation:
\begin{equation}
n/N(T=0)=\exp(-E_s/k_BT)
\end{equation}
where $n$ is the adlayer occupation number at a temperature $T$,
and $N(T=0)$ is the number of atoms in a layer at zero
temperature. Eq. (7) holds as long as the surface structure is
preserved, and interactions between defects can be neglected.
Hence, we used the adlayer occupation data in a temperature region
in which the concentration of adatom-vacancy pairs is small. The
natural logarithm of adlayer occupation vs. $1/k_bT$ is shown in
Fig.~\ref{adlocc} for the V(111), V(001) and V(011) surfaces, respectively.
 The calculated surface defect formation energies of
the various low-index faces of vanadium are tabulated in
Table~\ref{tab}.

An alternative method of estimating values of $E_s $ is to use the
contribution of defects to anharmonicity, as reflected in the
temperature dependence of the mean square vibration amplitude
$<u^2> $. Because the lattice in the vicinity of a defect is
distorted, the value of $<u^2> $ will be larger for atoms near a
defect. The layer averaged contribution to  $<u^2> $ will
therefore be proportional to the number of defects $n$, and hence
to $\exp(-E_s/k_BT)$. To extract $E_s $, we have fitted our $<u^2>
$ data to the form
\begin{equation}
<u^2>= aT + b(T) \exp(-E_s/k_BT)
\end{equation}
where $a$ is a constant, and $b(T)$ is a low order polynomial in
$T$. We found that taking $b(T)$ to be first order in $T$, namely
$b(T)=bT$, was sufficient to obtain a good fit. Typical fits were
shown as solid lines in Figs. \ref{uxz_111} and \ref{u2}.
Values of $E_s $ deduced from these fits for the various surfaces
are given in Table~\ref{tab}. There is good agreement between the
values of $E_s $ obtained by the two methods.

To summarize this subsection, the formation energy of surface defects
is lowest at the V(111) surface, and increases at the V(001) and
V(011) surfaces. This hierarchy is consistent with the results
obtained for various faces of copper
 (fcc lattice) by H$\ddot a$kkinen et al.\cite{Hakkinen}
In both cases, the surface defect formation energy is largest for
the close packed surfaces (V(011) in case of a bcc lattice, and
Cu(111) in case of a fcc lattice), and lowest for the least packed
surfaces,  V(111) and Cu(011), respectively.

\begin{table}
\caption{\label{tab} Formation energy, $E_s$, of adatom-vacancy
pairs. Top line-Eq.(7), second line-Eq.(8). The data for Cu is
from Ref.~\cite{Hakkinen} Units are in eV. }
%\begin{ruledtabular}
\begin{tabular}{|c|c|c|c|} \hline
Surface & $(111)$ & $(001)$ & $(011)$ \\ \hline \hline
 $Va~(bcc)~$ & $0.6 \pm 0.2~eV $ & $0.67 \pm 0.03~eV $ & $2.68 \pm 0.2~eV $ \\ \hline
$Va~(bcc)~$ & $0.53 \pm 0.05~eV $ & $0.73 \pm 0.05~eV $ & $2.5 \pm
0.3~eV $ \\ \hline
 $Cu~(fcc)~E_s$ & $1.92~eV $ & $0.86~eV $ & $0.39~eV $ \\ \hline
\end{tabular}
%\end{ruledtabular}
\end{table}

\subsection{\label{sec:level1} Diffusion coefficients }

Mass transport at surfaces can be characterized using the self
diffusion coefficients. These coefficients are found from the
particle trajectories, $\vec r_{i,\mu}(t)$, by calculating the
average mean square displacement $ R_{l,\mu}^2$
\begin{equation}
 R_{l,\mu}^2=\left< \frac{1}{n_l} \sum_{i \in l } [ \vec r_{i,\mu}(t+\tau) - \vec r_{i,\mu}(\tau)]^2 \right >
\end{equation}
where $\mu=x,y,z$ is a coordinate index, the sum includes atoms in
the layer $l$, and the angular brackets denote averaging over time
from the origin ($\tau$).
\begin{figure}[h]
\centerline{\epsfxsize=10.0cm \epsfbox{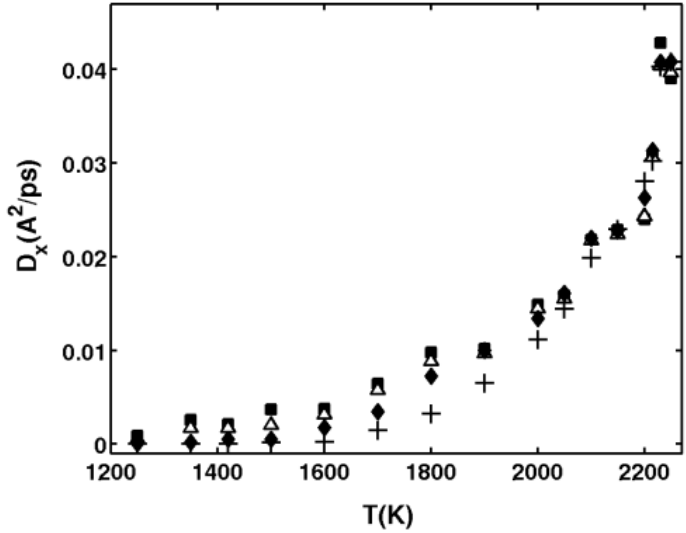}}
\caption{\label{d2230K} Diffusion coefficients in layers, vs
temperature, for the V(111) system along the x-direction
($[1\bar{1}0]$).
 Squares correspond to the adlayer, $l=0$, triangles to the first layer, $l=1$, diamonds to the second layer, $l=2$
 and crosses denote coefficients of diffusion in layer $l=3$. }
\end{figure}

The diffusion coefficients $ D_{l,\mu}$ are calculated separately
for each layer in the $x,y$ and $z$ directions, according to the
Einstein relation
\begin{equation}
 D_{l,\mu}=\lim_{t \rightarrow \infty } \frac{R^2_{l,\mu}}{2t}
\end{equation}
Values of the diffusion coefficients versus layer number are shown in
Fig.~\ref{d2230K}. As expected, the mobility of atoms is larger
the closer one comes to the surface. At
high enough temperatures, where the surface layers become
quasiliquid, the diffusion coefficient is the same (within the
error bars) for all the atoms in the region of the surface. These
observations correlate with the structural variations in the
surface region exhibited in the pair correlation functions, the
structure order parameters, and local density profiles.

\begin{figure}[h]
\centerline{\epsfxsize=10.0cm \epsfbox{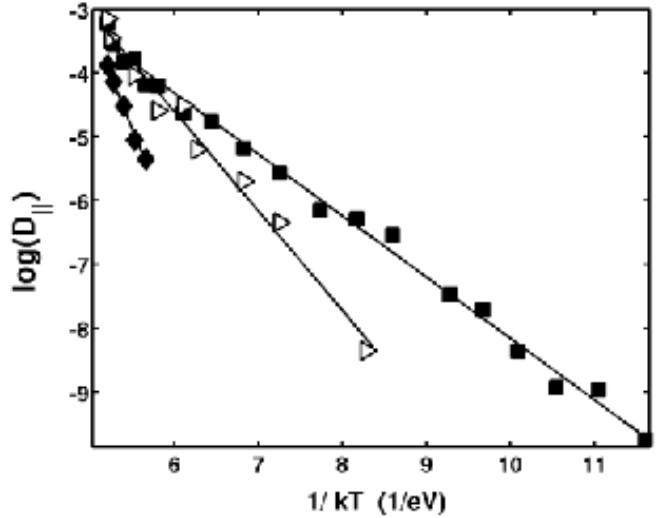}}
\caption{\label{diffuse} Plot of natural logarithm of the in-plane
diffusion coefficient $\log (D_{||})$, for the various faces of
vanadium  as a function of inverse temperature. $(D_{||})$ is in units of $\rm \AA ^{2}/psec$.}
\end{figure}

The natural logarithm of the in-plane diffusion coefficients,
$D_{||}=\frac{1}{2}(D_x+D_y)$, for the first surface layer of the
various faces vs.  temperature is shown in Fig.~\ref{diffuse}.
Within our accuracy, the diffusion can be characterized by a
simple exponential dependence throughout the whole temperature
range, $D(T)\propto \exp{(-E_d/k_BT)}$, with diffusion activation
energy, $E_d$,  being independent of temperature.
 Values of $E_d$  for the various low-index faces
are given in Table~\ref{tab1}. The diffusion coefficient of the
least packed V(111) face is the largest. The diffusion coefficient
of V(001) is smaller, but still larger than the diffusion
coefficient of the close packed V(011) surface. In the temperature
range where the surface region is still solid-like, diffusion
takes place mainly by atoms exchanging places with vacancies.
Writing $E_d$ as a sum of the formation energy, $E_s $, and
migration energy, $E_m $, and using the values of $E_s $ from the
preceding section, we can obtain values of $E_m $ for the various
surfaces. These are listed in Table~\ref{tab1}. It is seen that the
migration energy for the V(111) surface is again the lowest, while
for the other two surfaces $E_m $ is the same within the error
bars of our calculation.

\begin{table}
\caption{\label{tab1} Calculated diffusion activation energy $E_d
$ and migration energy $E_m $ }
\begin{tabular}{|c|c|c|c|} \hline
Surface & $V(111)$ & $V(001)$ & $V(011)$ \\ \hline \hline
 $E_d $(eV) & $0.96 \pm 0.02~eV $ & $1.56 \pm 0.03~eV $ & $3.29 \pm 0.2~eV $ \\ \hline
 $E_m $(eV) & $0.43 \pm 0.07~eV $ & $0.87 \pm 0.08~eV $ & $0.79 \pm 0.5~eV $ \\ \hline
\end{tabular}
\end{table}

\section{\label{sec:level1} Conclusions }

We investigated premelting effects at the various surfaces of
vanadium using molecular dynamics simulations. We determined the
thermodynamic melting temperature of vanadium described by the FS
potential as $2220 \pm 10$ K ($V(111)$ surface), and studied
structural, transport (diffusion) and energetic properties
(formation energy of surface defects).

We found that the surface region of $V(111)$ begins to disorder
first, via the generation of vacancy-adatom pairs and a formation
of an adlayer at temperatures above $1000~K$. At higher
temperatures, the surface region becomes quasiliquid. This process
begins above  $1600~K$ for the V(001) surface. At the closest
packed V(011) surface, this effect is observed only in close
proximity of $T_m$.

The results of our simulations of surface premelting of the bcc
metal, vanadium, are similar to the results obtained for various
fcc metals, in the sense that the onset of disorder is seen first
at the surface with the lowest density.

\section{\label{sec:level1} Acknowledgments }

We are  grateful to  Dr. G. Wagner and A. Kanigel for their contribution to this project.
The authors would like to acknowledge the use of computer resources
belonging to the High Performance Computing Unit,
a division of the Inter University Computing Center,
which is a consortium formed by research universities in Israel.
More information about this facility can be found at $http://www.hpcu.ac.il.$
This study was supported in part by the
Israeli Science Foundation, the German-Israeli Foundation, and by
the Technion VPR Fund for Promotion of Research.

\end{document}